\documentstyle[12pt]{article}
\begin{document}
\vspace{2cm}

\begin{center}
{\bf\large 
SO(4) symmetry  
and off-diagonal long-range order in the Hubbard bilayer}
\end{center}

Short title: Hubbard bilayer

\begin{center}
 Yu Shi
\\Fudan-T. D. Lee Physics Laboratory and Department of Physics,
\\Fudan University, Shanghai 200433, 
People's Republic of China
\end{center}

\newpage
\begin{abstract}
Yang's $\eta$ pairing operator is generalized to explore off-diagonal 
long-range order in the Hubbard bilayer with an arbitrary 
chemical potential. With this operator and a constraint condition on 
annihilation and creation operators, we construct explicitly
 eigenstates which possess simultaneously three 
kinds of off-diagonal long-range order, i.e., the intralayer one and the
interlayer one for on-site pairing, and that for interlayer nearest-neighbor
pairing.  As in the simple Hubbard model there is also an SO(4) symmetry,
with the  the generators  properly defined.
A sufficient condition leads to at least one of the above three kinds of 
off-diagonal long-range order. A constraint relation among different  
kinds of off-diagonal long-range order is also given.
There exists a triplet of collective modes if 
the U(1) symmetry of a subgroup is spontaneously broken.
\end{abstract}

\newpage
\section*{I. Introduction}
  Off-diagonal long-range order (ODLRO)  is essential for the phenomena
  of superconductivity and superfluidity \cite{yang}. The BCS wave function
  does have ODLRO via cooper pairing \cite{bardeen}, however, it is not an
  eigenstate of the Hamiltonian. In the search for the mechanism of high
  temperature superconductivity, the Hubbard model is studied intensively. 
  Yang constructed explicitly many
  eigenstates, which is metastable for attractive interaction, 
  of the Hubbard Hamiltonian
  possessing ODLRO \cite{yang1}.  Furthermore,
  Yang and Zhang uncovered an SO(4) symmetry and put forward a sufficient  
  condition for any state to possess ODLRO \cite{yang2}. 
  Then Zhang predicted a triplet of collective modes, 
  the massless Goldstone mode and a pair of massive modes, in the
  superconducting state \cite{zhang1}.

  In this paper we extend these discussions to the Hubbard bilayer, i.e., 
  two Hubbard planes coupled by interlayer hopping. The motivation is
  that strong coupling between the adjacent {\rm $CuO_{2}$} planes of one
  double plane has been directly revealed experimently \cite{tranquada}
  \cite{rossat}\cite{mook}\cite{stern}.
  The $\eta$ pairing operator generalized from Yang's is introduced in 
  Sec. II. Explicit eigenstates with ODLRO are constructed in Sec. III. 
  SO(4) symmetry and the sufficient condition for a state
  to possess ODLRO is discussed in Sec. IV, where a constraint relation
  among different kinds of ODLRO is also given. In 
  Sec. V, the triplet of collective modes are discussed.
  A brief summary is contained in Sec. VI, 
  where some  open problems are raised.

\section*{II. $\eta$ pairing}
The Hamiltonian is 
\begin{eqnarray}
H&=&T_{\parallel}(A)\,+\,T_{\parallel}(B)\,+\,T_{\perp}\,+\,V,\\
T_{\parallel}(A)&=&t_{\parallel}\displaystyle\sum_{r, \delta}
(a_{r\uparrow}^{\dagger}a_{r+\delta\uparrow}
+a_{r\downarrow}^{\dagger}a_{r+\delta\downarrow}),\nonumber\\
&=&\displaystyle\sum_{k}\epsilon(k)(a_{k\uparrow}^{\dagger}a_{k\uparrow}+
a_{k\downarrow}^{\dagger}a_{k\downarrow}),\\
T_{\parallel}(B)&=&t_{\parallel}\displaystyle\sum_{r, \delta}
(b_{r\uparrow}^{\dagger}b_{r+\delta\uparrow}
+b_{r\downarrow}^{\dagger}b_{r+\delta\downarrow})\nonumber\\
&=&\displaystyle\sum_{k}\epsilon(k)(b_{k\uparrow}^{\dagger}b_{k\uparrow}+
b_{k\downarrow}^{\dagger}b_{k\downarrow}),\\
T_{\perp}&=&t_{\perp}\displaystyle\sum_{r}
(a_{r\uparrow}^{\dagger}b_{r\uparrow}
+a_{r\downarrow}^{\dagger}b_{r\downarrow}
+b_{r\uparrow}^{\dagger}a_{r\uparrow}
+b_{r\downarrow}^{\dagger}a_{r\downarrow})\nonumber\\
&=&t_{\perp}\displaystyle\sum_{k}
(e^{-iq_{z}d}a_{k\uparrow}^{\dagger}b_{k\uparrow}+
e^{-iq_{z}d}a_{k\downarrow}^{\dagger}b_{k\downarrow}+
e^{iq_{z}d}b_{k\uparrow}^{\dagger}a_{k\uparrow}+
e^{iq_{z}d}b_{k\downarrow}^{\dagger}a_{k\downarrow}),\\
V&=&U\displaystyle\sum_{r}
[(a_{r\uparrow}^{\dagger}a_{r\uparrow}-\mu)
(a_{r\downarrow}^{\dagger}a_{r\downarrow}-\mu)
+(b_{r\uparrow}^{\dagger}b_{r\uparrow}-\mu)
(b_{r\downarrow}^{\dagger}b_{r\downarrow}-\mu)],
\end{eqnarray}
where 
$r$ is a two-dimensional integral coordinate variable on each $L\times L$ 
lattice, $\delta$ is its nearest-neighbors on the plane. $k$ is  
a two-dimensional integral momentum variable.  
$q_{z}$ denotes the momentum perpendicular to  the planes, the distance 
between which is $d\,=\,z_{A}\,-\,z_{B}$, where $z_{A}$ and $z_{B}$ are
the vertical coordinates of the planes A and B, respectively.
$a_{r\uparrow}$ and $a_{r\downarrow}$ are the coordinate-space annihilation 
operators for spin-up and spin-down electrons on  layer A, respectively,
$b_{r\uparrow}$ and $b_{r\downarrow}$ are those on layer B.  
$a_{k\uparrow}$, $a_{k\downarrow}$, $b_{k\uparrow}$ and $b_{k\downarrow}$
are the corresponding annihilation operators in the momentum space.
$\mu U$ is the chemical potential.
$T_{\parallel}(A)$ and $T_{\parallel}(B)$ are kinetic energies on the two 
layers, respectively. $T_{\perp}$ is the interlayer coupling. V is the
on-site electron-electron interaction potential energy. 
Following Ref. \cite{yang1},  we use
\begin{equation}
\epsilon(k)\,=\,4-2\cos k_{x}-2\cos k_{y} \label{eq:posi}
\end{equation}
to make the $T_{\parallel}(A)$ and $T_{\parallel}(B)$ positive.
Actually as has been pointed out \cite{yang1}, only 
$\epsilon(k)+\epsilon(\pi-k)\,=$ constant is required to make
$\eta$ pairing possible.

For our present model, the operator $\eta$ is introduced as
\begin{equation}
\eta\,=\,\eta_{\parallel}(A)+\eta_{\parallel}(B)+\eta_{\perp},
\end{equation}
with
\begin{eqnarray}
\eta_{\parallel}(A)&=&\displaystyle\sum_{r}e^{-i\pi\cdot r}
a_{r\uparrow}a_{r\downarrow}\nonumber\\
&=&e^{2iq_{z}z_{A}}\displaystyle\sum_{k}a_{k\uparrow}a_{\pi-k\downarrow},
\end{eqnarray}
\begin{eqnarray}
\eta_{\parallel}(B)&=&\displaystyle\sum_{r}e^{-i\pi\cdot r}
b_{r\uparrow}b_{r\downarrow}\nonumber\\
&=&e^{2iq_{z}z_{B}}\displaystyle\sum_{k}b_{k\uparrow}b_{\pi-k\downarrow},
\end{eqnarray}
and
\begin{eqnarray}
\eta_{\perp}\,&=&\,\displaystyle\sum_{r}e^{-i\pi\cdot r}
(a_{r\uparrow}b_{r\downarrow}-a_{r\downarrow}b_{r\uparrow})\nonumber\\
&=&e^{iq_{z}(z_{A}+z_{B})}
\displaystyle\sum_{k}(a_{k\uparrow}b_{\pi-k \downarrow}
-a_{k\downarrow}b_{\pi-k\uparrow}),
\end{eqnarray}
where $\pi$ is two-dimensional. Actually $\eta_{\parallel}(A)$ and
$\eta_{\parallel}(B)$ are on-site $\pi$-momentum pairings, 
while $\eta_{\perp}$ is the interlayer nearest-neighbor $\pi$-momentum
pairing. The definition is, of course, only meaningful when $L$ is even. 

One can obtain the commutators
\begin{eqnarray}
[T_{\parallel}(A)+T_{\parallel}(B),\, 
\eta_{\parallel}(A)+\eta_{\parallel}(B)]&=&
-8t_{\parallel}[\eta_{\parallel}(A)+\eta_{\parallel}(B)], 
\end{eqnarray}
\begin{eqnarray}
[T_{\parallel}(A)+T_{\parallel}(B),\, \eta_{\perp}] 
&=&-8t_{\parallel}\eta_{\perp}, 
\end{eqnarray}
\begin{eqnarray}
[T_{\perp},\,\eta_{\parallel}(A)+\eta_{\parallel}(B)]
&=&-2t_{\perp}\eta_{\perp},
\end{eqnarray}
\begin{eqnarray}
[T_{\perp},\,\eta_{\perp}]
&=&-2t_{\perp}[\eta_{\parallel}(A)+\eta_{\parallel}(B)].
\end{eqnarray}
Under the constraint condition
\begin{equation}
a_{r\uparrow}^{\dagger}a_{r\uparrow}+
b_{r\downarrow}^{\dagger}b_{r\downarrow}\,=\,
a_{r\downarrow}^{\dagger}a_{r\downarrow}+
b_{r\uparrow}^{\dagger}b_{r\uparrow}\,=\,1,  \label{eq:cons}
\end{equation}
we may obtain
\begin{eqnarray}
[V,\,\eta_{\parallel}(A)+\eta_{\parallel}(B)]&=&-(1-2\mu)U
(\eta_{\parallel}(A)+\eta_{\parallel}(B)), 
\end{eqnarray}
\begin{eqnarray}
[V,\,\eta_{\perp}]&=&-(1-2\mu)U\eta_{\perp}.
\end{eqnarray}

Therefore under the constraint condition (\ref{eq:cons}), we have 
\begin{equation}
[H,\,\eta]\,=\,-E\eta,               \label{eq:hc}
\end{equation}
where $E\,=\,8t_{\parallel}+2t_{\perp}+(1-2\mu)U$.

The total momentum operator $P$ is
\begin{equation}
P\,=\,\displaystyle\sum_{k}({\bf k}+{\bf q_{z}})
(a_{k\uparrow}^{\dagger}a_{k\uparrow}
+a_{k\downarrow}^{\dagger}a_{k\downarrow}+b_{k\uparrow}^{\dagger}b_{k\uparrow}
+b_{k\downarrow}^{\dagger}b_{k\downarrow}),
\end{equation}
the commutator of which with $\eta$ is
\begin{equation}
[P,\,\eta]\,=\,-\pi \eta.  \label{eq:pc}
\end{equation}

\section*{III. Eigenstates with ODLRO}
We can generate the state 
\begin{equation}
\psi_{N}\,=\,\beta(\eta^{\dagger})^{N}|vac>, \label{eq:psi}
\end{equation}
where $|vac>$ is the vacuum state, 
\begin{equation}
\beta\,=\,(N!(M-N)!/M!)^{1/2},\, M\,=\,L^{2} 
\end{equation}
is the normalization factor.  
Because of (\ref{eq:hc}) and (\ref{eq:pc}), 
$\psi_{N}$ is a simultaneous eigenstate  of $H$ and $P$,
\begin{eqnarray}
H\psi_{N}&=&NE\psi_{N},\\
P\psi_{N}&=&N\pi\psi_{N}.
\end{eqnarray}

Just similar to Ref. \cite{yang1}, one can construct another 
state $\psi'_{N}$, by replacing $\eta$ in Eq. (\ref{eq:psi}) with
the 0-momentum pairing operator 
\begin{equation}
\eta_{0}\,=\,\displaystyle\sum_{r}(a_{r\uparrow}a_{r\downarrow}
+b_{r\uparrow}b_{r\downarrow}+a_{r\uparrow}b_{r\downarrow}
-a_{r\downarrow}b_{r\uparrow}).
\end{equation}
$\psi'_{N}$ has the same expectation value for $H$ as $\psi_{N}$ but is not 
an eigenstate. So $\psi_{N}$ is not the ground state. Similar to 
Ref. \cite{yang1}, it can be proved to be metastable if $U\,<\,0$. 

Now we examine ODLRO. By definition, a 
state $\psi$ possess ODLRO for a local operator 
$Q_{r}$ means that 
the ODLRO correlation function
\begin{equation}
\displaystyle\lim_{|r-s|\rightarrow\infty}\psi^{\dagger}Q_{s}^{\dagger}
Q_{r}\psi\,\neq\,0.
\end{equation}

In our system, there are three kinds of ODLRO correlation functions, 
the intralayer one for 
on-site pairing 
$\psi_{N}^{\dagger}a_{s\downarrow}^{\dagger}a_{s\uparrow}^{\dagger} 
a_{r\uparrow}a_{r\downarrow}\psi_{N}\,=\,$ 
$\psi_{N}^{\dagger}b_{s\downarrow}^{\dagger}b_{s\uparrow}^{\dagger} 
b_{r\uparrow}b_{r\downarrow}\psi_{N}$, 
the interlayer one for on-site pairing
$\psi_{N}^{\dagger}b_{s\downarrow}^{\dagger}b_{s\uparrow}^{\dagger} 
a_{r\uparrow}a_{r\downarrow}\psi_{N}\,=\,$ 
$\psi_{N}^{\dagger}a_{s\downarrow}^{\dagger}a_{s\uparrow}^{\dagger} 
b_{r\uparrow}b_{r\downarrow}\psi_{N}$, 
and that for interlayer pairing
$\psi_{N}^{\dagger}(b_{s\downarrow}^{\dagger}a_{s\uparrow}^{\dagger}
-b_{s\uparrow}^{\dagger}a_{s\downarrow}^{\dagger}) 
(a_{r\uparrow}b_{r\downarrow}-a_{r\downarrow}b_{r\uparrow})\psi_{N}$, 
$r\,\neq\,s$.
They all equal to
\begin{equation}
\frac{N(M-N)}{M(M-1)}e^{i\pi\cdot(r-s)}.
\end{equation}
Therefore three kinds of ODLRO are  simultaneously possessed by $\psi_{N}$.

\section*{IV. SO(4) symmetry}
Define $\frac{\eta^{\dagger}}{2}\,=\,J_{+}\,=\,J_{x}+iJ_{y}$, 
$\frac{\eta}{2}\,=\,J_{-}\,=\,J_{x}-iJ_{y}$,
$J_{z}\,=\,
 \frac{1}{4}\displaystyle\sum_{k}
(a_{k\uparrow}^{\dagger}a_{k\uparrow}
+a_{k\downarrow}^{\dagger}a_{k\downarrow}
+b_{k\uparrow}^{\dagger}b_{k\uparrow}
+b_{k\downarrow}^{\dagger}b_{k\downarrow}
+a_{k\uparrow}^{\dagger}b_{k\uparrow}
+a_{k\downarrow}^{\dagger}b_{k\downarrow}
+b_{k\uparrow}^{\dagger}a_{k\uparrow}
+b_{k\downarrow}^{\dagger}a_{k\downarrow})
-\frac{M}{2}$.
One may find that $J_{x}$, $J_{y}$ and $J_{z}$ are generators of an
SU(2) symmetry. Under the constraint condition Eq. (\ref{eq:cons}), 
\begin{equation}
[H, J^{2}]\,=\,[H, \,J_{z}]\,=\,0.
\end{equation}

If  the total momentum operator is trivially re-defined as
\begin{equation}
P'\,=\,\sum_{k}({\bf k}-\frac{1}{2}{\bf \pi}+{\bf q_{z}})
(a_{k\uparrow}^{\dagger}a_{k\uparrow}
+a_{k\downarrow}^{\dagger}a_{k\downarrow}
+b_{k\uparrow}^{\dagger}b_{k\uparrow}
+b_{k\downarrow}^{\dagger}b_{k\downarrow}),
\end{equation}
we have 
\begin{equation}
[P',\,{\bf J}]\,=\,0.
\end{equation}
From (\ref{eq:hc}) we know that different from simple Hubbard model
where $[H,\,{\bf J}]\,=\,0$ \cite{yang1}, we 
cannot have $[H,\,{\bf J}]\,=\,0$ for coupled bilayer 
even if we re-define $\epsilon(k)$
as $-2\cos k_{x}-2\cos k_{y}$  and let $\mu\,=\,1/2$, since 
$[T_{\perp},\,\eta]$ cannot vanish.

The particle-hole pairing operator is              
\begin{equation}
\zeta\,=\,\displaystyle
\sum_{r}(a_{r\uparrow}a_{r\downarrow}^{\dagger}
+b_{r\uparrow}b_{r\downarrow}^{\dagger})
=\sum_{k}(a_{k\uparrow}a_{k\downarrow}^{\dagger}
+b_{k\uparrow}b_{k\downarrow}^{\dagger}).
\end{equation}
Defining $\zeta^{\dagger}\,=\,J_{x}'+iJ_{y}'$, $\zeta\,=\,J_{x}'-J_{y}'$,
$J_{z}'\,=\,\frac{1}{2}\displaystyle\sum_{k}(a_{k\downarrow}^{\dagger}a_{k\downarrow}
+b_{k\downarrow}^{\dagger}b_{k\downarrow}$$
-a_{k\uparrow}^{\dagger}a_{k\uparrow}-b_{k\uparrow}^{\dagger}b_{k\uparrow})$,
one may find that $J_{x}'$, $J_{y}'$, $J_{z}'$ are
generators of an SU(2) symmetry, 
 and that  
\begin{equation}
[H,\,{\bf J'}]\,=\,[P',\,{\bf J'}]\,=\,0. \label{eq:spin}
\end{equation}
Note that the constraint condition  Eq. (\ref{eq:cons}) is not required for
Eq. (\ref{eq:spin}), which just represents the SU(2) symmetry of spin.

Consequently, as in the simple Hubbard, there is also an 
SO(4) symmetry 
in the Hubbard bilayer, with the generators properly defined.
Most of the discussions in Ref. \cite{yang2} can thus generalized in 
a straightforward way, for example, many eigenfunctions for $H$ and $P'$ can
be obtained explicitely, the symmetry properties of energy spectrum 
are almost also valid. One only needs to take into account that now  
there are  two planes and  
the system is also symmetric under the exchange of the two planes. 

However, the sufficient condition for a state to possess ODLRO should be 
re-considered.
It can be stated as follows. 

{\em Theorem.} For any state $\psi$, if $j^{2}-j_{z}^{2}\,=\,O(M^{2})$, 
where $j$ and $j_{z}$ are the quantum numbers of $J^{2}$ and $J_{z}$,
respectively, then there is at least 
one kind of ODLRO among intralayer one for on-site pairing, 
interlayer one for on-site pairing and that for 
interlayer nearest-neighbor pairing.

{\em Proof.} Assume matrix $Q$ has matrix 
element which is the ODLRO correlation function  
of the operator $\Delta$,
\begin{equation}
Q_{sr}\,=\,<\Delta_{r}|Q|\Delta_{s}>\,=\,
<\psi|\Delta_{s}^{\dagger}\Delta_{r}|\psi>,
\end{equation}
Using
\begin{equation}
<\Delta_{s}|\phi>\,=\,M^{-1/2}e^{i\pi\cdot s}
\end{equation}
as the trial wavefunction for $Q$, we obtain the expectation value of
$Q$
\begin{equation}
<Q>\,=\,<\phi|Q|\phi>\,=\,\frac{1}{M}\sum e^{i\pi\cdot(s-r)}
<\psi|\Delta_{s}^{\dagger}\Delta_{r}|\psi>.
\end{equation}
On the other hand, we have
\begin{equation}
4(j^{2}-j_{z}^{2}+j+j_{z})\,=\,<\psi|\eta^{\dagger}\eta|\psi>\,=\,
\sum e^{i\pi\cdot(s-r)}(2Q^{(1)}_{sr}+2Q^{(2)}_{sr}+Q^{(3)}_{sr}),
\end{equation}
where 
\begin{eqnarray}
Q^{(1)}_{sr}\,&=&
\,<\psi|a_{s\downarrow}^{\dagger}a_{s\uparrow}^{\dagger}a_{r\uparrow}
a_{r\downarrow}|\psi>
\,=\,<\psi|b_{s\downarrow}^{\dagger}b_{s\uparrow}^{\dagger}
b_{r\uparrow}b_{r\downarrow}|\psi>\\
Q^{(2)}_{sr}\,&=&
\,<\psi|a_{s\downarrow}^{\dagger}a_{s\uparrow}^{\dagger}b_{r\uparrow}b_{r\downarrow}|\psi>
\,=\,<\psi|b_{s\downarrow}^{\dagger}b_{s\uparrow}^{\dagger}a_{r\uparrow}a_{r\downarrow}|\psi>\\
Q^{(3)}_{sr}\,&=&
\,<\psi|(b_{s\downarrow}^{\dagger}a_{s\uparrow}^{\dagger}-b_{s\uparrow}^{\dagger}a_{s\downarrow})
(a_{r\uparrow}b_{r\downarrow}-a_{r\downarrow}b_{r\uparrow})|\psi>.
\end{eqnarray}
Therefore 
\begin{equation}
2<Q^{(1)}>+2<Q^{(2)}>+<Q^{(3)}>\,=\,\frac{1}{M}(j^{2}-j_{z}^{2})+O(1).
\end{equation}
If $j^{2}-j_{z}^{2}\,=\,O(M^{2})$, 
at least the largest of the eigenvalues of $Q^{(1)}$, $Q^{(2)}$ and $Q^{(3)}$ 
is O(M),
according to Yang's Theorem \cite{yang},
there is corresponding ODLRO.
Q.E.D.

Tian proved a theorem that  the ODLRO correlation function of                                    
a local operator $R_{r}$ decays if there is another local operator
$S_{r}$ satisfying $[H, \,S_{r}]\,=\,R_{r}$ \cite{tian}.
Using this theorem, we can obtain a useful result giving constraint  
on the ODLRO. 
By calculating 
$[H,\,\displaystyle\sum_{r}a_{r\uparrow}a_{\downarrow}]$, we know that
$t_{\parallel}Q^{(nn)}+t_{\perp}Q^{(3)}-U(2\mu-1)Q^{(1)}$ decays, here
$Q^{(1)}$ and $Q^{(3)}$ are as defined above, 
$Q^{(nn)}$ is the ODLRO 
correlation function for intralayer nearest-neighbor 
s-wave pairing 
$\displaystyle\sum_{r,\delta}(a_{r\uparrow}a_{r+\delta\downarrow}-a_{r\downarrow}
a_{r+\delta\uparrow})$. 
So the existence of one of the three ODLRO implies at least one of the other two.
Furthermore, if $\mu\,=\,1/2$, interlayer pairing ODLRO exists if and only if
ODLRO for intralayer nearest-neighbor s-wave pairing exists.  
Consistent result has been obtained for the pairing amplitudes in the 
states with spontaneous gauge symmetry breaking \cite{shi}.

\section*{V. Collective modes}
In the Hubbard model, the SU(2) pseudospin symmetry contains the U(1)  
phase symmetry as a subgroup, whose spontaneously breaking, i.e., the 
superconductivity, gives rise to a pair of massive collective modes which 
together with the usual Goldstone mode form a triplet representation of the 
pseudospin symmetry \cite{zhang1}. Here we may see that these arguments 
can also extended  to the bilayer model. 

First it can be found that the operators $\Delta_{-}\,=\,\frac{1}{\sqrt{2}}
\displaystyle\sum_{r}(a_{r\uparrow}a_{r\downarrow}
+b_{r\uparrow}b_{r\downarrow}+a_{r\uparrow}b_{r\downarrow}
-a_{r\downarrow}b_{r\uparrow})$, 
$\Delta_{+}$ = $-\Delta_{-}^{\dagger}$, and 
$\Delta_{0}\,=\,\frac{1}{2}\displaystyle\sum_{r\sigma}
(a_{r\sigma}^{\dagger}a_{r\sigma}+b_{r\sigma}^{\dagger}b_{r\sigma}
+a_{r\sigma}^{\dagger}b_{r\sigma}+b_{r\sigma}^{\dagger}a_{r\sigma})$
form an irreducible tensor of rank 1 under the SU(2) defined by $J_{x}$,
$J_{y}$ and $J_{z}$ in the last section.  Then consider the response 
function 
\begin{equation}
D_{\alpha}(t,t')\,=\,
-\frac{i}{M}\theta(t-t')<[J_{\alpha}(t),\Delta_{\alpha}(t')]>,
\end{equation}
where $\alpha$ denotes +, - or 0.
One may obtain
\begin{equation}
D_{0}\,=\,\frac{\rho}{\omega+i\delta},
\end{equation}
\begin{equation}
D_{\pm}(\omega)\,=\,\frac{<\Delta_{\pm}>}{\omega \pm E+i\delta}.
\end{equation}

Therefore if the U(1) symmetry generated by $J_{z}$ is spontaneously broken,
i.e., if there is superconductivity with 
intralayer or(and) interlayer pairings,
$<\Delta_{+}>$ = $-<\Delta_{-}>^{*}\,\neq\,0$, 
then there is a triplet of collective modes with energies $\pm E$ and 0, 
respectively.

\section*{VI. Discussions}
To summarize, by introducing an $\eta$ operator generalized from Yang's
and under a constraint condition, 
the discussions concerning ODLRO and SO(4) symmetry 
for simple Hubbard model are
extended to Hubbard bilayer. 
In the explicitly constructed eigenstates, there
are simultaneously three kinds of ODLRO, i.e., the intralayer one for  
on-site pairing, the interlayer one for on-site pairing, and that for 
interlayer nearest-neighbor pairing. By properly defining the generators,
it is found that there is also an SO(4) symmetry.
The sufficient condition
for a state to possess ODLRO is that at least one of the three kinds of
ODLRO exists if the condition is satisfied. We also obtain a constraint 
relation among the ODLRO for interlayer nearest-neighbor pairing, the 
intralayer ODLRO for on-site pairing, and that for intralayer s-wave
nearest-neighbor pairing.

Though the explicit eigenstates possess the three kinds of ODLRO  
simultaneously, the sufficient condition can only ensure that at  least
one of them exsits. Another kind of ODLRO involves in the constraint 
relation. Further studies are needed to clarify whether these kinds of 
ODLRO must exist simultaneously.

Some exact results about the ground state of the simple Hubbard model are
based on the exchange of $\eta$ (pseudospin) and $\zeta$ (spin) operators
under particle-hole transformation of the spin-up electrons \cite{lieb}
\cite{shen}. This situation  is lost for the bilayer since the interlayer 
pairing is   involved in our $\eta$ operator.

Recently the collective modes predicted by Zhang \cite{zhang1} was 
generalized to the triplet case, which is then claimed to be responsible 
for the 41 meV peak in the neutron scattering spectrum on $YBa_{2}CuO_{x}$
\cite{zhang2}. However the dependence on $q_{z}$ \cite{tranquada}
could not be explained. Our results suggest that the bilayer model 
might be incorporated into the framework of this theory to account for 
the $q_{z}$ dependence. 

\section*{ACKNOWLEDGEMENT}
R.B. Tao is acknowledged for helpful discussions.

\end{document}